\documentstyle[12pt]{article}
\begin{document}
\begin{center}
{\Large \bf Comment on ``Quantum Entanglement and\\ the
Nonexistence of Superluminal Signals''} \\[1.5cm]
{\bf Vladimir S.~MASHKEVICH}\footnote {E-mail:
mash@gluk.apc.org}  \\[1.4cm]
{\it Institute of Physics, National academy
of sciences of Ukraine \\
252028 Kiev, Ukraine} \\[1.4cm]
\vskip 1cm

{\large \bf Abstract}
\end{center}

We argue that the claim given in quant-ph/9801014 is
untenable. The fallacy in the proof is a misinterpretation
of the no-cloning theorem, which does not allow quantum
jumps, specifically measurements.

\newpage

\section{What is the claim}

In a recent paper [1] Westmoreland and Schumacher  made an
attempt to show that ``ordinary quantum mechanics is not
consistent with the superluminal transmission of classical
information.'' They assert that ``quantum entanglement
can be used to show that superluminal classical communication
is impossible...''. The results relate to ``a universe in
which both quantum mechanics and relativity hold true.''

\section{What is the proof}

The ``proof is constructed from three elements: the
`no-cloning' theorem of quantum mechanics..., quantum
teleportation..., and the relativity of simultaneity.''

The time evolution of states involved is as follows:
\begin{equation}
\left| \phi_{C} \right\rangle\left| \Psi^{-}_{AB}
\right\rangle\stackrel{{\rm I}i}{\longrightarrow}
\left| \chi^{i}_{CA} \right\rangle\left| \chi^{i}_{B}
\right\rangle\stackrel{{\rm II}i}{\longrightarrow}
\left| \chi^{i}_{CA} \right\rangle\left| \phi_{B}
\right\rangle
\label{1}
\end{equation}
where I$i$ is a quantum jump caused by Alice's measurement
for $CA$, II$i$ is Bob's unitary transformation for $B$.

Due to a superluminal message transmitted from Alice to
Bob, there exists a frame of reference in which $CB$ system
is in the state
\begin{equation}
\left| \phi_{C} \right\rangle\left| \phi_{B} \right\rangle
\label{2}
\end{equation}
for some time. This, the authors conclude, contradicts the
no-cloning theorem, which completes the proof.

\section{What is wrong}

The fallacy in the proof is a misinterpretation of the
no-cloning theorem, which states:
\begin{equation}
(\forall(U,\left| 0_{Y},0_{M} \right\rangle))
(\exists\left| a_{X} \right\rangle)
(\forall\left| \psi_{M} \right\rangle):
U\left| a_{X},0_{Y},0_{M} \right\rangle\ne
\left| a_{X},a_{Y},\psi_{M} \right\rangle,
\label{3}
\end{equation}
or, equivalently,
\begin{equation}
(\neg\exists(U,\left| 0_{Y},0_{M} \right\rangle))
(\forall\left| a_{X} \right\rangle)
(\exists\left| \psi_{M} \right\rangle ):
U\left| a_{X},0_{Y},0_{M} \right\rangle=
\left| a_{X},a_{Y},\psi_{M} \right\rangle,
\label{4}
\end{equation}
where $U$ is a unitary operator on the Hilbert space $H_{XYM}$.
In words: There do not exist $U$ and $\left| 0_{Y},0_{M}
\right\rangle$, such that for every $\left| a_{X}
\right\rangle$ there exists $\left| \psi_{M} \right\rangle$,
such that
\begin{equation}
U\left| a_{X},0_{Y},0_{M} \right\rangle=
\left| a_{X},a_{Y},\psi_{M} \right\rangle
\label{5}
\end{equation}
holds.

The theorem does not allow quantum jumps, specifically
measurements, whereas (\ref{1}) involves a quantum jump.

\section{What is shown}

Eq.(\ref{1}) describes teleportation, so that the authors
showed, in effect, that superluminal communication
allows superluminal teleportation, which is trivial.

\section{What is the actual state of affairs}

The authors cite Wheeler: ``Can we find an argument against
using entanglement for superluminal communication that
is as simple and clear as the no-cloning theorem?'' There
exists such an argument in the limits of special relativity:
There is no preferred quantum-jump hypersurface (or pair
of causally separated events) [2]. (But in a complete
dynamical description of quantum jumps given in the series
of our papers [3], this argument fails.)

\end{document}